\documentclass[smallextended]{svjour3}       % onecolumn (second format)
\smartqed  % flush right qed marks, e.g. at end of proof
\usepackage{graphicx}

\usepackage{hyperref}

\usepackage{booktabs}
\usepackage{amssymb}
\usepackage{amsmath}
\usepackage{colortbl}
\usepackage{rotating}
\usepackage{xspace}

\def\N{{\rm I\kern-.5ex N}}
\def\Z{{\sf \vrule height 1.55ex depth-1.2ex width.03em\kern-.11em Z%
        \kern-.9ex Z\kern-.11em\vrule height 0.3ex depth0ex
width.03em}}
\def\Q{{\rm\kern.2ex\vrule height1.55ex depth-.05ex width.03em\kern-
.7ex Q}}
\def\R{{\rm I\kern-.5ex R}}
\def\C{{\rm\kern.3ex\vrule height1.55ex depth-.05ex width.03em\kern-
.7ex C}}

\def\MS{\mathit{MS}}
\def\MScore{\mathrm{MScore}}
\def\SGC{\mathit{SGC}}
\def\barx{\overline{\textbf{x}}}
\DeclareMathOperator*{\argmin}{arg\,min}
\begin{document}

\title{A Realistic Model under which the Genetic Code is Optimal}

\authorrunning{Buhrman et al.} % if too long for running head

\author{Harry Buhrman \and Peter T.\ S.\ van der Gulik \and Gunnar W.\ Klau \and 
Christian Schaffner \and Dave Speijer \and Leen Stougie}

\institute{H.~Buhrman \and P.~T.~S.~van der Gulik \and G.~W.~Klau \and 
C.~Schaffner \and L.~Stougie\at
              Centrum Wiskunde \& Informatica (CWI) \\
              P.O. Box 94079\\
              1090 GB Amsterdam, the Netherlands\\
              \email{peter.van.der.gulik@cwi.nl}
           \and
           D.~Speijer \at
           Academic Medical Center\\
           Department of Medical Biochemistry\\
	   Meibergdreef 15\\
           1105 AZ Amsterdam, the Netherlands
           \and
           H.~Buhrman \and C.~Schaffner \and D.~Speijer \at
           University of Amsterdam
           \and
           G.~W.~Klau \and L.~Stougie \at
           VU University Amsterdam
}

\date{}

\maketitle

\begin{abstract}
The genetic code has a high level of error robustness. Using values of hydrophobicity 
scales as a proxy for amino acid character, and the Mean Square measure as a function 
quantifying error robustness, a value can be obtained for a genetic code which reflects 
the error robustness of that code. By comparing this value with a distribution of values 
belonging to codes generated by random permutations of amino acid assignments, the level 
of error robustness of a genetic code can be quantified. We present a calculation in which 
the standard genetic code is shown to be optimal. We obtain this result by (1) using recently 
updated values of polar requirement as input; (2) fixing seven assignments (Ile, Trp, His, Phe, 
Tyr, Arg, and Leu) based on aptamer considerations; and (3) using known biosynthetic relations 
of the 20 amino acids.  This last point is reflected in an approach of subdivision (restricting 
the random reallocation of assignments to amino acid subgroups, the set of 20 being divided in 
four such subgroups). The three approaches to explain robustness of the code (specific selection 
for robustness, amino acid-RNA interactions leading to assignments, or a slow growth process of 
assignment patterns) are reexamined in light of our findings. We offer a comprehensive hypothesis, 
stressing the importance of biosynthetic relations, with the code evolving from an early stage 
with just glycine and alanine, via intermediate stages, towards 64 codons carrying todays 
meaning.

\end{abstract}

% ===========================================================================
\section{Introduction} \label{sec:introduction}
% ===========================================================================

The genetic code is a basic feature of molecular biology. It sets the rules according to which 
nucleic-acid sequences are translated into amino-acid sequences. The genetic code probably evolved by a 
process of gradual evolution from a proto-biological stage, via many intermediary stages, to its present 
form (see e.g. ~\cite{Crick68,LehmanJ88,vetsigianetal06}). During this process, error robustness was built into the code (see 
e.g.~\cite{ardell98,vetsigianetal06,higgs09,Crick68,IkeharaOAH02,FreelandWK03,CaporasoYK05,Wong05,WolfK07,Massey08,digiulio08}). 
Two different kinds of error robustness 
can be observed~\cite{vetsigianetal06} by even the most superficial inspection of the Standard Genetic Code (SGC). On 
one hand, codons assigned to the same amino acid are almost always similar, see Table~\ref{tab:tab1}. As an example, all 
codons ending with a pyrimidine (U or C) in a codon box (the four codons sharing first and second 
nucleotides) are without exception assigned to the same amino acid (e.g.\ UAU and UAC both code for 
Tyr). On the other hand, similar codons are mostly assigned to similar amino acids, e.g.\ codons 
with U in the second position are all assigned to hydrophobic amino acids~\cite{woese65,woeseetal66a,woeseetal66b}. 
This is illustrated in Table~\ref{tab:tab1}, when looking at the values of polar requirement: overall, low values 
of polar requirement correspond to hydrophobic amino acids. 

\begin{table}[h]
\setlength{\tabcolsep}{1.4mm}
\centering
  \begin{tabular}{ | ll | ll | ll | ll | }
    \hline
UUU &Phe (4.5) & UCU &Ser (7.5) & UAU &Tyr (7.7) & UGU &Cys (4.3) \\
UUC &Phe (4.5) & UCC &Ser (7.5) & UAC &Tyr (7.7) & UGC &Cys (4.3) \\
UUA &Leu (4.4) & UCA &Ser (7.5) & UAA &STOP      & UGA &STOP \\
UUG &Leu (4.4) & UCG &Ser (7.5) & UAG &STOP      & UGG &Trp (4.9) \\ \hline

CUU &Leu (4.4) & CCU &Pro (6.1) & CAU &His (7.9) & CGU &Arg (8.6) \\
CUC &Leu (4.4) & CCC &Pro (6.1) & CAC &His (7.9) & CGC &Arg (8.6) \\
CUA &Leu (4.4) & CCA &Pro (6.1) & CAA &Gln (8.9) & CGA &Arg (8.6) \\
CUG &Leu (4.4) & CCG &Pro (6.1) & CAG &Gln (8.9) & CGG &Arg (8.6) \\ \hline

AUU &Ile (5.0) & ACU &Thr (6.2) & AAU &Asn (9.6) & AGU &Ser (7.5) \\
AUC &Ile (5.0) & ACC &Thr (6.2) & AAC &Asn (9.6) & AGC &Ser (7.5) \\
AUA &Ile (5.0) & ACA &Thr (6.2) & AAA &Lys (10.2)& AGA &Arg (8.6) \\
AUG &Met (5.0) & ACG &Thr (6.2) & AAG &Lys (10.2)& AGG &Arg (8.6) \\ \hline

GUU &Val (6.2) & GCU &Ala (6.5) & GAU &Asp (12.2) & GGU &Gly (9.0) \\
GUC &Val (6.2) & GCC &Ala (6.5) & GAC &Asp (12.2) & GGC &Gly (9.0) \\
GUA &Val (6.2) & GCA &Ala (6.5) & GAA &Glu (13.6) & GGA &Gly (9.0) \\
GUG &Val (6.2) & GCG &Ala (6.5) & GAG &Glu (13.6) & GGG &Gly (9.0) \\
    \hline
  \end{tabular}
\caption{The standard genetic code. Assignment of the 64 possible
codons to amino acids or stop signals, with updated polar
requirement~\cite{mathewlutheyschulten08} values indicated in brackets. \label{tab:tab1} }
\end{table}
\noindent

Three main approaches exist to explain the emergence of this robustness of the code: specific selection 
for robustness (see e.g.~\cite{haighurst91,freelandhurst98a,vetsigianetal06}), amino acid-RNA interactions 
leading to assignments (see e.g.~\cite{woese65,yarusetal09}), and a slow growth process of 
assignment patterns reflecting the history of amino acid repertoire growth (see e.g.~\cite{Crick68,wong75,Massey06,digiulio08}).
The concept that all three competing hypotheses are important has also been brought forward~\cite{KnightFL99}. In 
the present study we make adjustments to earlier mathematical work in this field (see 
e.g.~\cite{haighurst91,freelandhurst98a,buhrmanetal11}) 
which integrate the three concepts into a single mathematical model. We will now, one by one, introduce 
these three adjustments.

\subsection{Polar Requirement}
\label{subsec:polarreq}

The polar requirement~\cite{woeseetal66a} is not just a measure related to hydrophobicity. Several different measures of 
hydrophobicity exist, each focusing on different aspects of it. Polar requirement specifically focuses on the nature of 
the interaction between amino acids and nucleic acids. Stacking interactions between 
e.g. the planar guanidinium group of arginine and the planar purine ring systems and pyrimidine ring systems of RNA 
is an example of that. Woese chose to chemically model the nucleotide rings by using pyridine as the solvent system in 
the measurements leading to the polar requirement scale~\cite{woese65,woeseetal66a,woeseetal66b,Woese67,Woese73}. 
This interaction between amino acids and nucleic acids has been stressed as an especially important aspect of early protein 
chemistry because one possibility for the very first function of coded peptides was suggested~\cite{Noller04} to be the enlargement of 
the number of conformations accessible for RNA (realized by the binding of small, oligopeptide cofactors). Thus polar requirement could 
have been among the most important aspects of an amino acid during early stages of genetic code evolution.

The remarkable character of polar requirement as a measure of amino acids in connection to the genetic code was found again 
and again throughout the years. Firstly, Woese found that distinct amino acids coded by codons differing only in the 
third position are very close in polar requirement, despite differences in general character~\cite{woeseetal66b}. The pair cysteine 
and tryptophan 
nicely exemplifies this. Secondly, Haig and Hurst~\cite{haighurst91} discovered that polar requirement showed the SGC to be 
special to a much larger degree than another scale of hydrophobicity (the hydropathy scale of Kyte and Doolittle~\cite{kytedoolittle82}). 
Thirdly, when 
Mathew and Luthey-Schulten updated the values of polar requirement~\cite{mathewlutheyschulten08} by \emph{in silico} methods (the most 
important 
change was believed to be due to a cellulose-tyrosine interaction artefact in the original experiments), the SGC showed a further 
factor 10 increase~\cite{butleretal09} in 
error robustness calculations. In all these developments the expectation that polar requirement would behave in a special 
way, as interaction between nucleotides and amino acids is biochemically important, was more than borne out by the results. One of 
the adjustments 
we introduce in our work compared to our earlier calculations~\cite{buhrmanetal11} is 
that in the present work we use the new, updated values of polar requirement (see Table~\ref{tab:tab1}).

\subsection{Aptamers}
\label{subsec:aptamers}

Oligonucleic-acid molecules that bind to a specific target molecule (e.g.\ a specific amino acid) are called 
aptamers~\cite{ellingtonszostak90}. Over the last two decades, many results have been obtained regarding specific 
binding of amino acids by RNA aptamers, mainly by Yarus and co-workers~\cite{majerfeldyarus94,illangasekareyarus02,yarusetal09}. 
For several amino acids, codons and anticodons were found in binding sites, in quantities higher 
than would be expected to occur by chance~\cite{yarusetal09}. In Table~\ref{tab:tab2}, a list of occurrences of 
anticodons in binding sites of RNA sequences is given, together with the articles in which these sequences were 
reported. Please note that the definition of anticodons used in these articles is: triplets complementary to 
codons. These anticodons are 
therefore not necessarily identical to the triplets found in tRNA molecules which 
are normally meant with the word `anticodon'. As an example: the triplet AUG is considered as an His anticodon 
because it is complementary to the His codon CAU. In tRNAs, however, the anticodon recognizing CAU is GUG 
(see~\cite{JohanssonEHBB08,GrosjeanCM10} for reviews on codon-anticodon interaction). We summarize published details on the 
aptamers for seven amino acids, and subsequently formulate a conclusion regarding the implications of the 
existence of these molecules for genetic-code error-robustness calculations. This conclusion is based on reasoning 
presented by the Yarus group concerning the existence of specific relationships between certain triplets 
and certain amino acids. These relationships could have led to evolutionary conserved assignments of these amino 
acids to these triplets, e.g.\ by a mechanism as presented in~\cite{yarusetal09}. 
\medskip

For Ile, Trp, and His, three binding motifs were described, respectively named the `UAUU-motif'~\cite{lozuponeetal03}, 
the `CYA-motif'~\cite{majerfeldyarus05,majerfeldetal10}, and the `histidine-motif'~\cite{majerfeldetal05}. As can 
be seen from the names, the anticodons UAU for Ile, and CCA for Trp, are characteristic for the motifs (`CYA' stands for 
`CUA or CCA'). In the case of His, both GUG and AUG (the anticodons for the two His codons CAC and CAU) are found in 
quantities higher than would be expected by chance~\cite{majerfeldetal05}.

\begin{table}[h]
\centering
  \begin{tabular}{ | l | l | l | }
    \hline
Amino Acid & Anticodon & References  \\ \hline
Ile & UAU & \cite[pages 415-419]{yarusetal09}  \\
Trp & CCA & \cite[page 1918]{majerfeldetal10}  \\
His & GUG, AUG & \cite[pages 413-414]{yarusetal09}  \\
Phe & GAA, AAA & \cite[page 420]{yarusetal09}  \\
Tyr & GUA, AUA & \cite[page 423]{yarusetal09}  \\
Arg & CCU, UCU, ACG, GCG, UCG, CCG & \cite[page 2]{janasetal10}  \\
Leu & CAA, GAG, UAG & \cite[page 420]{yarusetal09}  \\ \hline
  \end{tabular}
\caption{The occurence of anticodons in binding sites of the RNA sequences of amino-acid binding aptamers, and 
the references in which the actual RNA sequences can be found. \label{tab:tab2} }
\end{table}
\noindent

Although binding sites for Phe and Tyr have so far not been studied as
extensively as those for Ile, Trp, and His, the analysis of Yarus et al. ~\cite{yarusetal09} 
shows that the anticodons (GAA and AAA
for Phe, and GUA and AUA for Tyr) are present in the binding sites 
more often than would be expected on a random basis.

Both the CCU anticodon~\cite{janasetal10} and the UCG anticodon~\cite{yarusetal09} are 
present in Arg binding sites more often than would be expected on a random basis. 
Thus, a physico-chemical background was observed, compatible with: (1) Arg having more than 4 codons,
and (2) all 6 Arg codons sharing the same middle nucleotide.

A similar observation can be made for Leu, the other amino acid which is
encoded by six codons all having the same middle nucleotide. For this amino acid, however, only a single RNA sequence was
found binding the amino acid with specificity~\cite{yarusetal09}. 
Inspection of this sequence shows anticodons UAG, GAG, and CAA to 
be present in its binding parts.

Taking the combined results of Yarus and co-workers into consideration, we propose to fix 
assignments of Ile, Trp, His, Phe, Tyr, Arg, and Leu for calculations using random variants 
of the SGC.
\medskip

\subsection{Gradual Growth}
\label{subsec:growth}

In Section~\ref{sec:methods} we present our approach in detail. We use
Haig and Hurst's `mean square' measure, (as first proposed in~\cite{haighurst91}) to 
quantify the error robustness of a given code. With this measure, 
a relatively error-robust code gets a low value when compared to 
the average value of a large set of codes produced by random 
allocation of amino-acid assignments (see~\cite{buhrmanetal11} for 
a more in-depth treatment of the approach). The space of codes allowed 
to exist by the allocation procedure can be large (in the original work 
of Haig and Hurst~\cite{haighurst91} the space has a size of exactly 20! 
codes, which is $\approx2.433 \cdot 10^{18}$ codes). We call a code optimal if 
it reaches the minimum in error robustness calculations among all possible 
codes in a particular setting.
\medskip

In 1975, Wong proposed the coevolution theory of the genetic code~\cite{wong75}. According to this proposal, SGC codons 
assigned to an amino acid biosynthetically derived from another amino acid, were originally assigned to that `precursor' 
amino acid. As an example: Pro is biosynthetically derived from Glu. According to coevolution theory, the four Pro codons (CCN) 
would have originally encoded Glu. Without embracing all details of the original coevolution theory, or modern refinements of 
the theory~\cite{wong07,digiulio08}, something remarkable can be noted as a result of this way of looking at the SGC. Shikimate-derived 
amino-acids (Phe, Tyr, and Trp) all have U in the first position of the codon (Phe: UUY; Tyr: UAY; and Trp: UGG). Glu-derived 
amino-acids (Pro, Gln, and Arg) almost always have C in the first position of the codon (Pro: CCN; Gln: CAR, which stands for `CAA or 
CAG'; and Arg: AGR and CGN, where N stands for all 4 nucleotides). Asp-derived amino-acids (Ile, Met, Thr, Asn, and Lys) all have A 
in the first position of the codon (Ile: AUY and AUA; Met: AUG; Thr: ACN; Asn: AAY; and Lys: AAR). Codons with G in the 
first position all code for amino acids produced in Urey-Miller experiments\footnote{For a recent update on prebiotic synthesis 
  see~\cite{parkeretal11} and references therein.} (Val: GUN; Ala: GCN; Asp: GAY; Glu: GAR; and Gly: GGN). This `layered structure' of
the SGC was first pointed out explicitly by Taylor and Coates~\cite{taylorcoates89}. It may indeed suggest a sequential development of 
the repertoire of amino acids specified in the developing code, and a possibly sequential introduction of use of G, A, C, and U 
as first nucleotide in codons. The `layered structure' of the SGC is a regularity different from the well-known error-robust 
distribution of polar requirement~\cite{haighurst91}, which is pronounced in the first and the third, but not in the second position of 
the codon (please note: having, as a group, all the same nucleotide in the \emph{first} position, gives error robustness for the 
group character to changes in the \emph{second} and \emph{third} position). As is shown in Appendix~\ref{app:matrix}, it is possible 
to prove the presence of the `layered structure' quantitatively, when the appropiate set of values is developed and used as input.

Freeland and Hurst~\cite{freelandhurst98b} followed the concept of Taylor and Coates, and formally divided the 20 amino acids in 
four groups of five amino acids each: Gly, Ala, Asp, Glu, and Val in a first group which could be called `the prebiotic group'; 
a second group of amino acids with codons starting with A (Ile, Met, Thr, Asn, and Lys); a third group with codons mainly starting 
with C (Leu, Pro, His, Gln, and Arg); and, finally, a group with codons mainly starting with U (Phe, Ser, Tyr, Cys, and Trp). 
Division of the set of twenty in these four subsets was subsequently incorporated in the calculations on code error 
robustness~\cite{freelandhurst98b}. This approach reduced the size of the space from which codes could be sampled randomly in a drastic way: 
from a size of about $2 \cdot 10^{18}$ codes (see above) to a size of $(5!)^{4}$ codes (which is exactly $2.0736 \cdot 10^{8}$ codes). 
This space was called the `historically reasonable' set of possible codes~\cite{freelandhurst98b}. By 
sampling from the historically reasonable set of possible codes, we incorporate in the current study the notion of 
a chronologically-determined, layered structure of the SGC. 
\medskip

\subsection{Integration of assumptions}
\label{subsec:synthesis}

We have found that if: (1) the updated values for polar requirement are used as amino-acid attributes; 
(2) the assignments of seven amino-acids to codons are fixed following the rationale given above; and 
(3) the subdivision leading to the historically reasonable set of possible codes is used to define the 
space of code variations (which is also reduced in size by (2)), then the SGC is optimal. It is important to 
note that the constraints applied drastically reduce the size of the space: with applying both (2) and 
(3), the ``realistic space'' has a size of 11520 codes.

% ===========================================================================
\section{Methods} \label{sec:methods}
% ===========================================================================

We use the mean-square method developed by Allf-Steinberger~\cite{alffsteinberger69}, Wong~\cite{wong80}, Di Giulio~\cite{digiulio89}, and Haig and
Hurst~\cite{haighurst91}. For the mathematical formulation, we follow the approach of~\cite{buhrmanetal11} and consider the undirected graph
$G=(V,E)$ that has the 61 codons\footnote{In the original calculation,
  Haig and Hurst ignored the three ``stop codons'' encoding
  chain termination.} as its vertices and an edge between any two codons if they differ in only one position, yielding 263 edges. A code
$F$ maps each codon $c$ to exactly one amino acid $F(c)$. We denote by $r_{F(c)}$ the polar requirement of the amino acid that codon $c$
encodes in the code $F$ and by $\mathbf{r}$ the full vector of 20 values. The mean-square error function of code $F$ is then given by 
\begin{eqnarray*}
  \MS_0^{\alpha, \mathbf{r}}(F)= \frac{1}{N} \sum_{\{ c,c'\} \in E}
  \alpha_{c,c'}\left( r_{F(c)}-r_{F(c')} \right) ^2 \, \label{eq:MS}
\end{eqnarray*}
where the $\alpha_{c,c'}$ are the weights of the different mutations that can occur (corresponding to edges of the graph) 
and $N=\sum_{\{c,c'\} \in E} \alpha_{c,c'}$ is the total weight. Following Haig and Hurst~\cite{haighurst91}, we use a subscript 
0 to indicate the overall measure. If we set all 263 weights $\alpha_{c,c'}$ to 1, we get the original function described 
by~\cite{haighurst91} which we simply denote by $\MS_0(F)$. We also consider the following set of weights introduced by Freeland 
and Hurst~\cite{freelandhurst98a} which differentiates between transition errors (i.e.~U to C, C to U, A to G, G to A) and transversion 
errors and the position where they occur in the codon:
\begin{itemize}
\item $\alpha_{c,c'}=0.5$ if $(c,c')$ is a transversion in the first position or a transition in the second position,
\item $\alpha_{c,c'}=0.1$ if $(c,c')$ is a transversion in the second position,
\item $\alpha_{c,c'}=1$ otherwise.
\end{itemize}
Using weights for different codon positions implies the existence of a tRNA with a triplet anticodon during the process of code evolution. 
As we consider a process of gradual expansion of the repertoire of amino acids during the evolution of the SGC (see e.g.~\cite{Crick68,LehmanJ88,ardell98}) as the 
most likely mechanism -with duplication of tRNA genes, and subsequent divergence (cf.~\cite{Ohno70}) of their sequences and functions- we think this assumption 
is acceptable. This assumption does not necessarily imply the existence of 
protein aminoacyl-tRNA synthetases during all or part of the process 
of code evolution, as there could originally have been ribozymes which fulfilled their function. The value of error-robustness of a 
code $F$ using the set of weights introduced above will be denoted by $\MS_0^{\mathit{FH}}(F)$.

In principle, there are at least three ways in which one can improve the model of~\cite{haighurst91} to reflect biological
reality more accurately. The first possibility is to change how the level of error robustness is measured, e.g.\ by introducing weighting factors as described
above. Variations of the weighting factors used in the calculation show an even higher error robustness of the SGC, as noticed by
e.g.\ \cite{freelandhurst98a,gilisetal01,butleretal09}. The rationale behind changing weighting factors is improved reflection of 
natural selection pressures. It is, however, difficult to decide which weighting factors adequately reflect the natural selection pressures
operating during the early evolution of the genetic code (see comment 4 of Ardell in~\cite{novozhilovetal07} and the exchange of thoughts with 
respect to `column 4' in~\cite{higgs09}).

The second way to improve the model is to change the set of values representing amino-acid properties used as input in the
error-robustness calculation. For instance, one can use the values of hydropathy from~\cite{kytedoolittle82}, or the matrix of~
\cite{gilisetal01} instead of the polar requirement scale. In our paper, we use the values of the 2008 update of polar requirement 
by \emph{in silico} methods~\cite{mathewlutheyschulten08} given in Table~\ref{tab:tab1}. Work concerning the issue what an `ideal' 
set of twenty values would look like, and work considering different known sets of amino-acid properties is presented in 
appendices~\ref{app:eppstein} and~\ref{app:aaindex}.

The third way to improve the model is to change the size of the space from which random codes are sampled~\cite{buhrmanetal11}. The
incentive to enlarge that space (as was done in~\cite{buhrmanetal11}) is the wish to work from a space that encompasses all possible codes, or at 
least, all known codes. As indicated in~\cite{buhrmanetal11}, larger spaces are increasingly difficult to work with. The frequency distributions 
obtained by sampling from the larger spaces in~\cite{buhrmanetal11} highly coincide with the frequency distribution obtained from the original space 
(as presented in~\cite{haighurst91}). From this viewpoint, working in the original space is acceptable as a simplification.  In the current
study, we \emph{shrink} the size of the space, based on considerations of fixed assignments of certain codons, and combining this with the constraint 
of the historically reasonable set of possible codes of~\cite{freelandhurst98b}, as outlined in Section~\ref{sec:introduction}. 

\medskip

MATLAB-programs were used for the error-robustness calculations
and visualizations. All software can be found as supplemental information, 
or downloaded from~\url{https://github.com/cschaffner/gcode}.

% ===========================================================================
\section{Results} \label{sec:results}
% ===========================================================================

Among all genetic codes (in this particular setting of the problem), the SGC is optimal in terms of
error-robustness if:
\begin{enumerate}
\item We use the updated values of polar requirement~\cite{mathewlutheyschulten08}.
\item We use fixation for Phe, Tyr, Trp, His, Leu, Ile, and Arg, based on aptamer experiments~\cite{yarusetal09,janasetal10}. 
\item We use the historically reasonable set of possible codes~\cite{freelandhurst98b}.
\end{enumerate}

\begin{figure}[h]
  \includegraphics[width=0.85\linewidth]{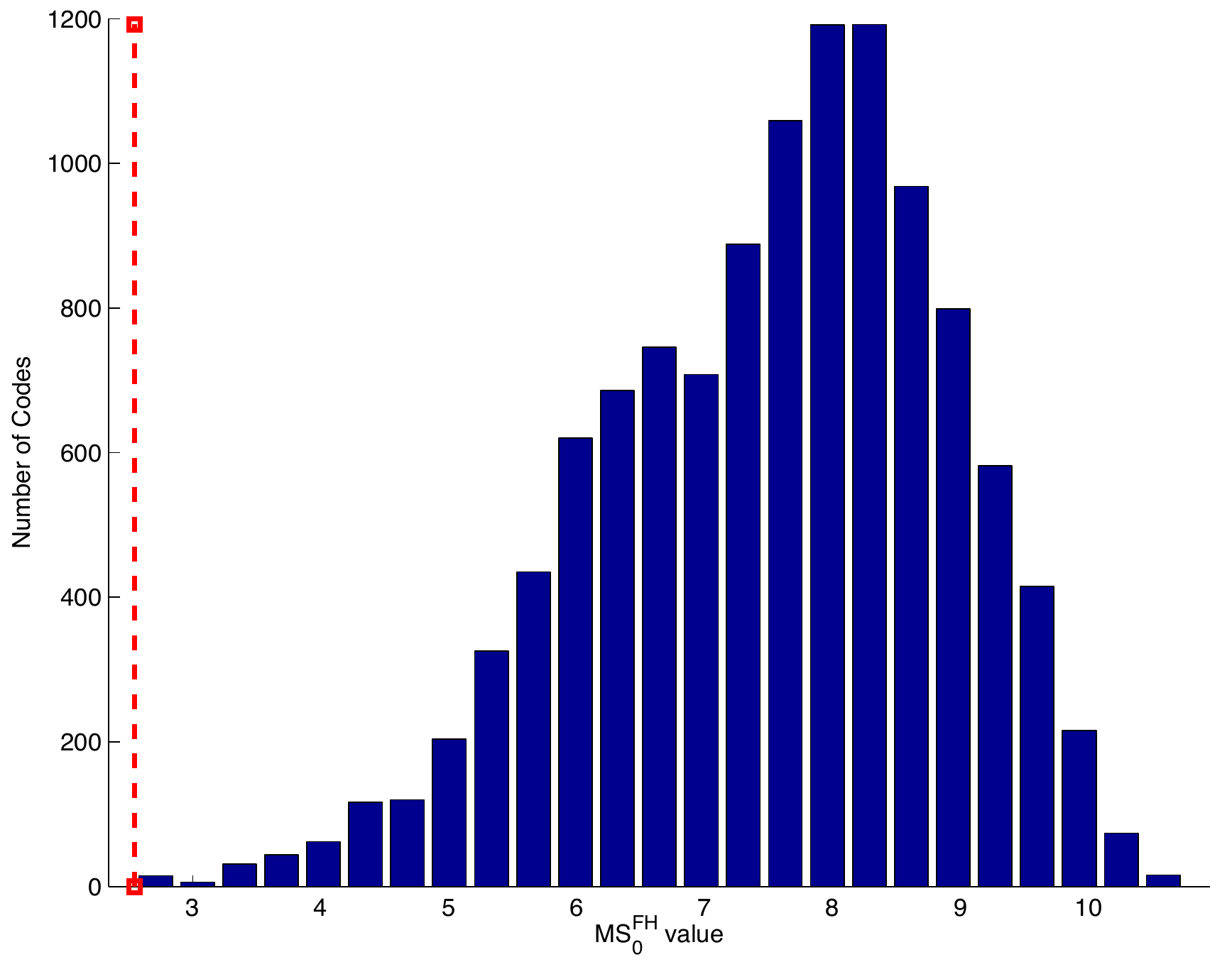}
\caption{Histogram of $\MS_0^{\mathit{FH}}$-values when using the historically reasonable set of possible codes, and fixing Phe, Tyr,
  Trp, His, Leu, Ile, Arg. Standard genetic code (indicated by dashed red line) is optimal.
\label{fig:sgcoptimal}}
\end{figure}

Figure~\ref{fig:sgcoptimal} shows a histogram of
$\MS_0^{\mathit{FH}}(F)$-values resulting from this 
procedure. When, the original error function $\MS_0(F)$ from~\cite{haighurst91} is 
used, the result is essentially the same: the SGC is the optimal 
code. We wondered if by fixation of just one or two more assignments, the SGC would be 
optimal in the space resulting from the combination of these fixations with the 
random permutations of amino acid assignments according to the method used by Haig and 
Hurst~\cite{haighurst91}, \emph{without} the constraint of the historically reasonable 
set of possible codes~\cite{freelandhurst98b}. This was not the case 
(as is reported in Appendix~\ref{app:minfixed}).

% ===========================================================================
\section{Discussion}  \label{sec:discussion}
% ===========================================================================

What is the biological relevance of the mathematical result presented, if any? Can we indeed 
conclude that natural selection steered the translation system toward better and better variants 
of the assignments (in terms of error-robustness) within realistic boundaries? Stated 
differently, when making a model, should one respect that seven assignments are \emph{fixed}, 
and that the system evolved \emph{gradually} (as reflected by using the historically reasonable
set of possible codes), until the optimal code (within these boundaries) was reached? Or is it rash 
to arrive at such a conclusion, and could one imagine positive selection for error-robustness to be 
an illusion?

The space of codes resulting from the constraints imposed on the calculations is a space of 
very limited size: only 11520 codes ($2!\cdot2!\cdot4!\cdot5!$). The fact that the SGC is 
optimal in this space is impressive, but of a different order of magnitude than the
near-optimalities in significantly larger spaces presented in earlier studies (e.g.\
\cite{freelandhurst98a,freelandetal00,gilisetal01,butleretal09,buhrmanetal11}). The impact of 
the different fixed assignments varies: for the $\MS_0$-values, it would theoretically suffice 
to fix the three assignments of Phe, Trp, and Arg (or any set containing them) in order to find 
the SGC to be optimal in the resulting space.\footnote{When using
  the Freeland and Hurst weights (and hence the $\MS_0^{\mathit{FH}}$-values),
  it is possible to fix another set of three amino-acids Phe, His, Trp
  in order to make the SGC optimal.} In this way, the SGC can be thought of as the global optimum 
in a space of $3!\cdot4!\cdot5!\cdot5!=2073600$ codes. We further refrain from presenting it thus, because 
in doing so we would abandon the physico-chemical facts which were the starting point for our 
calculations with fixed assignments.

It is also possible to \emph{increase} the number of fixed assignments (and in this way \emph{decrease} 
the size of the space of random code variants) even further.  A recent article~\cite{johnsonwang10} 
suggests that more than the seven assignments (listed in Table~\ref{tab:tab2}) are fixed.

The logical extreme of fixing assignments is that \emph{all} assignments of the SGC are fixed, as argued 
recently by Erives~\cite{erives11}. In his theory, a kind of RNA cage (pacRNA: proto-anti-codon RNA) is 
presented, in which different amino acids are bound by different kinds of `walls', which are exposing 
anticodons to the different amino acids. Although this model combines elegant explanations for several 
aspects of present-day tRNA functioning, it is very hard to get an objective measure for the specificity 
of amino acid-anticodon interactions in this model. In particular, the different possibilities allowed by 
`breathing' of the cage cast doubt on interaction specificity. Some objections can also be raised 
regarding the tRNA activation mechanism. Yarus and co-workers recently reported a very small ribozyme 
(only five nucleotides in length) which was experimentally shown to aminoacylate certain small RNAs using 
aminoacyl-NMPs as activated precursors~\cite{turketal10,yarus11}. Such an early activation mechanism, 
using NTPs as source of energy, is different from the one in Erives' model, where the $5^\prime$ end of the 
pacRNA is performing this role.
\medskip

Taking all considerations sketched above into account, it is possible to draw a tentative picture of 
genetic code evolution which is compatible with the indications concerning which aspects of code 
evolution are important. Code evolution probably followed classical mechanisms of gene duplication 
and subsequent diversification (here of `tRNA' genes and genes involved in aminoacylation). 
Evolution would be mainly by stop-to-sense reassignments~\cite{LehmanJ88}, with occasional reassignments 
in only slightly different new or developing uses of codons (cf.~\cite{ardell98,vetsigianetal06}), 
not yet massively present in protein-coding sequences (cf. the frozen accident 
concept~\cite{Crick68}). In a proto-biological stage, RNA would be absent while very small peptides could 
have been synthesized, e.g. by the Salt-Induced Peptide Formation (SIPF) reaction~\cite{SchwendingerR89,RodeSSB99}. 
Under prebiotic conditions especially Ala and Gly would be expected to be present in relatively large amounts 
(see e.g.~\cite{HiggsP09,PhilipF11}). Asp-containing peptides could possibly play a role in the 
origin of RNA, as they could position $\rm Mg^{2+}$ ions in the correct orientation to help 
polymerize nucleotides, and, concomitantly, keep these ions from stimulating RNA hydrolysis 
~\cite{Szostak12eightfold}. Asp content of peptides could be enriched in the presence of 
carboxyl-group binding montmorillonite surfaces~\cite{RodeSSB99}.

In the first stages of coded peptide synthesis, GCC and GGC probably were the only codons in mRNAs~\cite{EigenS78}, 
and coded peptides would consist of Ala and Gly. The remaining codons effectively would be stop 
codons~\cite{LehmanJ88}, although functioning without release factors: water would break bonds 
between tRNA and peptide whenever codons stayed unoccupied for too long. The `single-step biosynthetic distance' between 
Ala and pyruvate suggests a carbon storage role for these peptides; Gly allowing folding of such molecules. A mRNA/tRNA 
system functioning without a ribosome has been proposed by several 
authors~\cite{CrickBBW61,Woese73,LehmanJ88}. The first rRNA could then have been functioning in improved 
termination (see above). At this stage the proposal that coded peptides enlarge the possible 
range of RNA conformations should be taken into account~\cite{Noller04}.

In the next stage of \emph{coded} peptide synthesis, Asp and Val could have been added to the repertoire 
(see e.g.~\cite{EigenS78,ardell98,Ikehara02,higgs09,GulikMGBR09}). This would have been a crucial step: enabling 
\emph{directed} production of the important Asp-containing peptides~\cite{Szostak12eightfold,GulikMGBR09} as well as 
formation of something resembling protein structure, characterized by hydrophobic cores (Val) and 
hydrophilic exteriors (Asp). The emerging \emph{polypeptides} could have functioned in carbon storage, as 
mentioned above. Having started with trinucleotide codons, this aspect was retained, not because 
four nucleotide codons are in principle impossible, but this system allowed a further 
robust development (cf.~\cite{vetsigianetal06}). Depletion of prebiotic pools of either Ala, Gly, 
Asp, or Val (e.g. by excessive storage in coded peptides) could have led to the biosynthetic routes 
involving Gly, Ser, Val, Asp, Ala, and pyruvate. In this way the lack of an amino acid could in principle 
be resolved by use of the other three (cf. the hypothesized carbon storage function of coded peptides).

In a further stage, Ser, and Asp-derived amino acids like Asn and Thr would be added to the repertoire. 
Asn would be the first amino acid with an entirely biosynthetic origin (it is relatively 
unstable, and does not accumulate prebiotically). The production of Asn is known to be originally linked to enzymatic 
conversion of Asp to Asn on a tRNA (see e.g.\cite{wong07}). When instead of two molecules of pyruvate, one molecule 
of pyruvate and one molecule of alpha-keto-butyrate are fed into the Val biosynthesis pathway, Ile is produced instead. 
Therefore, when both Thr and Val biosynthesis are present, the evolution of just one enzyme (making alpha-keto-butyrate 
from Thr) suffices for the emergence of Ile. Aptamers can handle this amino acid, and these two factors (easy development from existing 
biochemistry and easy manipulation by RNA) could be responsible for the `choice' of Ile (cf.~\cite{PhilipF11}).

Larger amino acids like His and Gln would have appeared in a later stage of code development than Asp-derived 
amino acids like Asn and Thr. The reactions catalyzed by the few enzymes in the Leu biosynthesis which are not 
enzymes involved in Val biosynthesis (apart from leucine aminotransferase) 
are reminiscent of the first three reactions of the citric acid cycle~\cite{VoetV95}. Jensen~\cite{Jensen76} hypothesized 
that originally enzymes would have had much broader substrate specificity. With 
the citric acid cycle being `old', as well as important for bio-energetic reasons, and Val biosynthesis being present, the 
system could have produced an excess of Leu. Again, aptamers 
would be able to `handle' Leu. Existing biochemistry and aptamer potential would thus answer the 
question why Ile and Leu are part of the Set of Twenty, and e.g. norleucine and alpha-amino-butyric acid
are not (cf.~\cite{PhilipF11}). Linked to the citric acid cycle and important in nitrogen management are 
Glu and Gln. A  further expansion of the repertoire with a Glu-derived amino acid is 
the expansion with Arg. Two of the enzymes of the urea (nitrogen management) cycle are related to pyrimidine synthesis enzymes, 
two others to purine synthesis enzymes~\cite{BergTSureacycle}. The last enzyme in the cycle is arginase. 
This suggests an ancient accumulation of Arg as a side effect of RNA synthesis, upon Glu becoming a major cell component. 
Arginase could function in bringing the Arg concentration down to acceptable levels. Aptamers could also have evolved to manipulate Arg levels, allowing 
Arg to become part of the Set of Twenty. Again Jensen's concept of 
primordial broad substrate specificity~\cite{Jensen76} is essential to get a possible answer to the 
`Why these 20?' question: Arg could be part of the set, rather than ornithine and citrulline, because Arg 
accumulates, and Arg can be manipulated by aptamers.

In an advanced stage of code development aromatic amino acids would be added to the repertoire, and release 
factors would evolve. Van der Gulik and Hoff~\cite{GulikH11} have argued that codons UUA, AUA, UAA, CAA, AAA, 
GAA, UGA, and AGA could not function unambiguously until the anticodon modification machinery was developed, 
which is seen by them as the last development leading to the full genetic code. Because 
archaea and bacteria have different solutions for the `AUA problem' (agmatidinylation vs. lysidinylation~\cite{GulikH11}), 
unambiguous sense assignment of AUA must have been late indeed. 

The SGC has probably evolved in a genetic environment characterized by rampant horizontal 
gene-flow~\cite{vetsigianetal06}. The interaction between genetic systems with slightly different, still-evolving 
codes, is thought to have caused both universality and optimality of the SGC~\cite{vetsigianetal06}. 
Universality, because the genetic code functioned as an innovation sharing protocol~\cite{vetsigianetal06}. 
Optimality, because competition allowed selection for the ability to translate the genetic information 
accurately~\cite{vetsigianetal06}. The work presented in our paper illuminates constraints within 
which this process of genetic code development took place. Both the step-by-step increasing complexity of 
biochemistry, and the stereochemical relationship between at least some amino acids and triplets, are factors 
which have to be taken into account. 

In summary, although there are at least two different lines of research suggesting a greater number of fixed 
assignments than the seven given in Table~\ref{tab:tab2} (based on the work of Yarus and 
co-workers~\cite{yarusetal09,janasetal10}), for now it is not clear that more (or even all~\cite{erives11}) assignments 
are fixed. Thus, the observed error-robustness still needs explanation. It is possible that the optimality of the SGC we found results 
from positive selection for error-robustness, though starting within a more restricted set of possibilities than previously thought.

\begin{acknowledgements}
 We thank the EiC and two anonymous reviewers for suggestions which improved the manuscript. Part of this research 
 has been funded by NWO-VICI Grant
 639-023-302, by the NWO-CLS MEMESA Grant, by the Tinbergen Institute, and by a NWO-VENI Grant.
\end{acknowledgements}

%############################################################################
%          BIBLIOGRAPHY
%############################################################################

%\bibliographystyle{abbrv}
%\bibliographystyle{alpha}
\bibliographystyle{spmpsci}      % mathematics and physical sciences
\bibliography{HPGCDL}

\appendix
% ===========================================================================
\section{Appendices} \label{sec:appendices}
% ===========================================================================

Four further observations are reported here. Firstly, as explained in Section~
\ref{sec:introduction}, consideration of the biosynthetic pathways leading to the 
different amino acids suggests an aspect of organization of the SGC in which 
GNN codons tend to be assigned to `prebiotic amino acids', ANN codons to 
comparatively small, aspartate-derived amino acids, CNN codons to larger 
amino acids, and UNN codons to the largest, or (in the case of cysteine) 
the most instable and reactive amino acid. In other words: the \emph{first} 
position of the codon might have a link with the complexity of biochemistry, 
e.g.\ the UNN codons being the only ones encoding aromatic amino acids and 
the instable cysteine, and reflecting the most advanced stage of biochemistry 
during the evolution of the genetic code (when the biochemistry was sufficiently 
complex to handle cysteine, and to build tryptophan). In Appendix~\ref{app:matrix}, 
we study this link with the biosynthetic development of amino acids by measuring 
how many one-atom changes are required to transform one amino acid into another.
With respect to this distance measure, amino acids derived from the same precursor 
(like e.g.\ Ile and Thr) are comparatively close, because they share 
structure parts. Changing the \emph{second} position of the codon (in the case of 
Ile and Thr: changing AUU to ACU) would then replace an amino acid 
by one with a comparatively similar structure, reflecting their membership of the 
same biosynthetic family. If the error-robustness calculation is performed with 
these molecular-structure distances, the SGC is found to have error protection in 
substitution mutations in the \emph{second} position (and therefore grouping e.g.\ ANU 
codons together). The results are given in Appendix~\ref{app:matrix}.

Secondly, we tried to find numerical values for the 20 amino acids which
make the SGC optimal in terms of error robustness among all possible genetic 
codes. Using a numerical optimization approach developed by Eppstein~\cite{eppstein03}, 
we were able to find 20 such values. In fact, many different sets of 20 values 
have this property. Details about these SGC-optimality calculations can be 
found in Appendix~\ref{app:eppstein}.

Thirdly, we screened a large list of physico-chemical amino-acid characteristics on their performance 
in our error-robustness calculations. Polar requirement was one of the best performing measures.
This strongly supports the remark by Haig and Hurst (``The natural code 
is very conservative with respect to polar requirement. The striking 
correspondence between codon assignments and such a simple measure deserves 
further study.''~\cite{haighurst91}). The observation of Vetsigian, Woese, 
and Goldenfeld (``Although we do not know what defines amino-acid `similarity' in the case 
of the code, we do know one particular amino-acid measure that seems to express it quite remarkably 
in the coding context. That measure is amino-acid polar requirement [...]''~\cite{vetsigianetal06}) 
should also be mentioned. More details are given in Appendix~\ref{app:aaindex}.

Finally, we wondered if, by fixing just one or two more assignments, the SGC would be optimal 
without using the subdivision leading to the historically reasonable set of possible codes 
(as explained in Section~\ref{sec:introduction}). This was not the case. When working with 
Haig-Hurst weights (i.e. equal weighting), there exist 34 sets of 9 fixed assignments which 
do have this characteristic. However, none of these 34 sets consists of the seven fixed 
assignments based on aptamer considerations plus two more amino acids. The smallest set 
containing the seven has size 10. When working with Freeland-Hurst weights (see Section 
\ref{sec:methods}), sets of 8 or 9 fixed assignments with the required characteristic, do not 
exist. This work is presented in Appendix~\ref{app:minfixed}.

\begin{figure}[hbtp]
\centering   \includegraphics[width=0.9\linewidth]{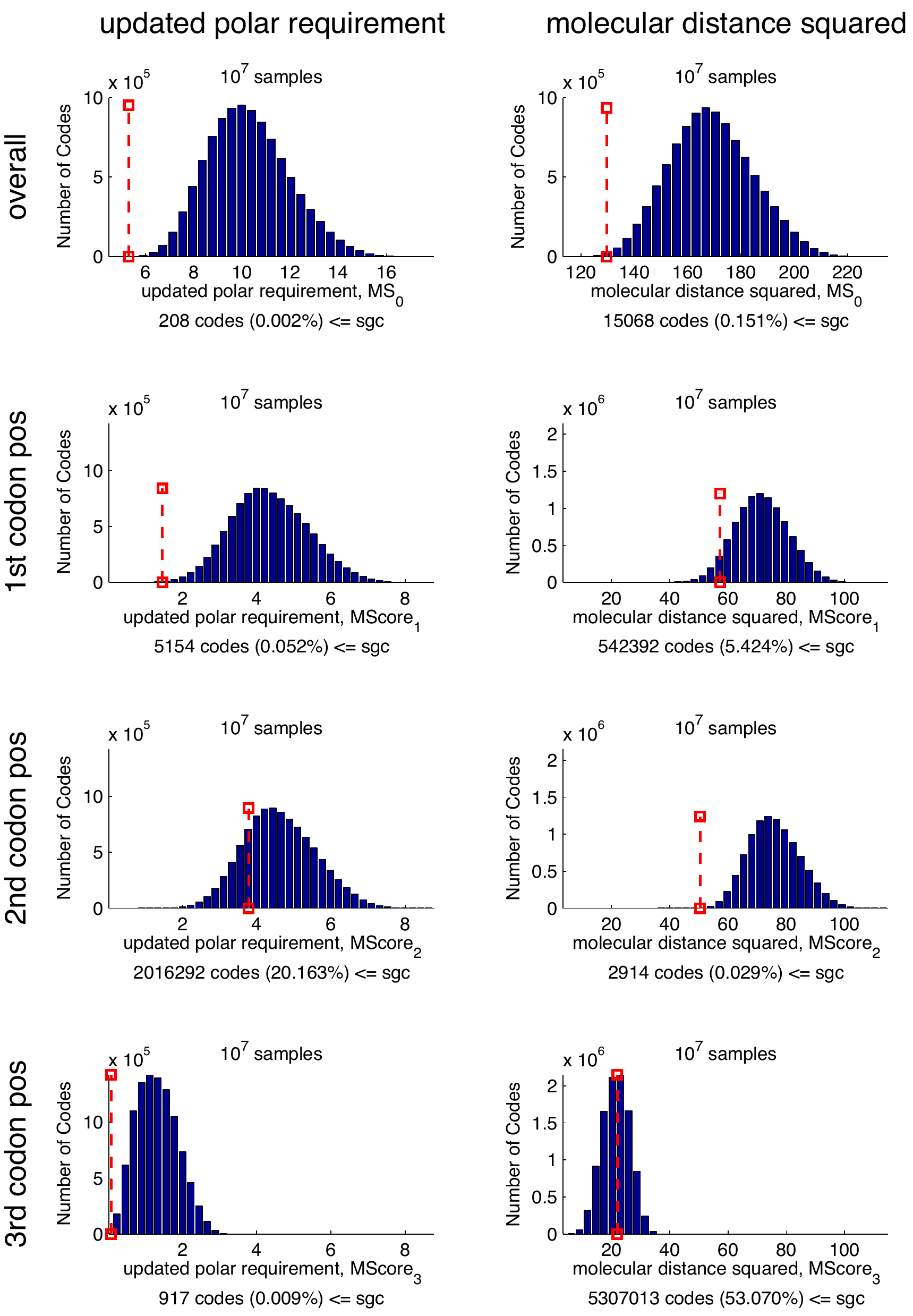}
\caption{Histograms of the MS-values of 10 million random samples using
  updated polar requirement~\cite{mathewlutheyschulten08} (4 histograms on the left) and
  molecular-structure distances from Table~\ref{tab:molstructure}
  squared (4 histograms on the right). The top row shows the $\MS_0$ value,
  the second row is the component from the first codon position
  ($\MScore_1$), third and forth row the components from the middle ($\MScore_2$) and last
  ($\MScore_3$) codon position. In contrast to the original definition~\cite{haighurst91}
  of $\MS_i$ for $i\geq 1$, we have chosen to normalize $\MScore_i$ with
  the same constant as $\MS_0$ so that $\MS_0 = \sum_{i=1}^3 \MScore_i$. The dashed red line indicates the value of
  the SGC. 
  \label{fig:splithistograms}}
\end{figure}

\subsection{Molecular Structure Matrix} \label{app:matrix}

Polar requirement is just one physico-chemical aspect of amino acids. The discovery that 
only 1 in 10000 random codes has a lower error-robustness value than the SGC when 
polar requirement is used as an amino-acid characteristic~\cite{haighurst91} 
is compelling evidence that error robustness is present in the SGC. 
When a conservative attitude is taken, and a phenomenon is considered noteworthy 
only when the probability to encounter it as a random effect is less than 
0.1 \%, the SGC is clearly noteworthy. If one considers the error-robustness 
values for the three positions separately (please refer to~\cite{buhrmanetal11} 
for details) the results in the left column of Figure~\ref{fig:splithistograms} 
are obtained. The third position is in the less than 0.1 \% category, the first 
position is in the less than 1 percent category, while the second position, 
with about 22 \%, is not even in the less than 5 \% category, and can thus 
not be considered special.

This result is not entirely satisfactory, because the codons of several pairs 
of similar amino acids are related by second position changes. For instance, a change 
from phenylalanine (Phe) to tyrosine (Tyr) is clearly a conservative 
change from a biological viewpoint. To develop a measure for this kind of 
amino-acid relatedness, we introduce a new way of measuring amino-acid
similarity by one-atom changes which yields a measure of similarity in
terms of molecular structure. We should stress that this measure does \emph{not} 
reflect actual chemical reactions/steps. As an example, we compute
the distance between Phe and Tyr to be 3 as follows: the hydrogen atom at 
the end of the side chain of Phe is taken off as a first step. An oxygen atom is 
placed on the position which the hydrogen atom had before as a second step. The 
Tyr molecule is completed by addition of an hydrogen atom on top of this oxygen 
atom, producing the hydroxyl group at the end of the side chain of Tyr, and this 
is the third and final step. 
Generally, the distance between two molecules is defined to be the
minimal number of ``allowed one-atom changes'' to transform one molecule into the other,
where the allowed one-atom changes are the following:
\begin{itemize}
\item taking off or attaching an arbitrary single atom,
\item creating or destroying a single bond (thereby possibly opening
  or closing a ring structure),
\item changing a single bond to a double bond or \emph{vice versa}.
\end{itemize}

It is not hard to see that an algorithmic way of computing the
distance between two molecules $m_1$ and $m_2$ is to find the maximal
common sub-graph $m_c$ of their molecular structure and to sum up how
many steps are required to go from $m_1$ to $m_c$ and from $m_2$ to
$m_c$. The distance matrix between the 20 amino acids in
Table~\ref{tab:molstructure} has been obtained in this way, using the
Small Molecule Subgraph Detector (SMSD) toolkit~\cite{rahmanetal09} to
find the maximal common subgraph and post-processing this information
with a python script. The software code can be found in the supplemental
information.

\newcommand{\rowcol}{\rowcolor[gray]{0.9}} 

\begin{table}[h]
\centering
\small
\setlength{\tabcolsep}{1.4mm}
  \begin{tabular}{ l|
rrr
>{\columncolor[gray]{.9}}r>{\columncolor[gray]{.9}}r>{\columncolor[gray]{.9}}r
rrr
>{\columncolor[gray]{.9}}r>{\columncolor[gray]{.9}}r>{\columncolor[gray]{.9}}r
rrr
>{\columncolor[gray]{.9}}r>{\columncolor[gray]{.9}}r>{\columncolor[gray]{.9}}r
rr }
Phe & 0\\
Leu & 15&  0\\ 
Ile & 21& 10&  0\\
\rowcol
Met & 21& 14& 14&  0\\
\rowcol
Val & 22& 15&  5& 11&  0\\
\rowcol
Ser & 17& 12& 14& 10& 11&  0\\ 
Pro & 17&  8&  8& 10& 11& 10&  0\\ 
Thr & 20& 13&  9&  9&  6&  5&  9&  0\\ 
Ala & 16& 11& 13&  9& 10&  3&  9&  8&  0\\
\rowcol
Tyr & 3& 16& 22& 22& 23& 18& 18& 21& 17&  0\\ 
\rowcol
His & 18& 15& 17& 17& 18& 13& 13& 16& 12& 19&  0\\ 
\rowcol
Gln & 20& 13& 13& 11& 12& 11&  9& 10& 10& 21& 12&  0\\ 
Asn & 19& 14& 16& 12& 13&  8& 12& 11&  7& 20& 13& 13&  0\\ 
Lys & 17& 12& 12& 14& 15& 14&  8& 13& 13& 18& 17& 13& 16&  0\\ 
Asp & 18& 13& 15& 11& 12&  7& 11& 10&  6& 19& 14& 12&  5& 15&  0\\ 
\rowcol
Glu & 19& 12& 12& 10& 11& 10&  8&  9&  9& 20& 15&  5& 12& 12& 11&  0\\ 
\rowcol
Cys & 17& 12& 14& 10& 11&  4& 10&  9&  3& 18& 13& 11&  8& 14&  7& 10&  0\\ 
\rowcol
Trp & 12& 23& 27& 27& 28& 23& 23& 26& 22& 15& 18& 22& 25& 23& 24& 25& 23&  0\\ 
Arg & 24& 15& 15& 17& 18& 17& 11& 16& 16& 25& 10& 12& 19& 15& 18& 15& 17& 24&  0\\ 
Gly & 19& 14& 14& 12& 11&  6& 12&  9&  5& 20& 15& 13& 10& 16&  9& 12&
6& 25& 19&  0\\ \hline 
\rule{0mm}{8.3mm}& \begin{sideways}Phe\end{sideways} 
& \begin{sideways}Leu\end{sideways}
& \begin{sideways}Ile\end{sideways}
& \begin{sideways}Met\end{sideways}
& \begin{sideways}Val\end{sideways}
& \begin{sideways}Ser\end{sideways}
& \begin{sideways}Pro\end{sideways}
& \begin{sideways}Thr\end{sideways} 
& \begin{sideways}Ala\end{sideways}
& \begin{sideways}Tyr\end{sideways}
& \begin{sideways}His\end{sideways}
& \begin{sideways}Gln\end{sideways}
& \begin{sideways}Asn\end{sideways}
& \begin{sideways}Lys\end{sideways}
& \begin{sideways}Asp\end{sideways} 
& \begin{sideways}Glu\end{sideways}
& \begin{sideways}Cys\end{sideways}
& \begin{sideways}Trp\end{sideways}
& \begin{sideways}Arg\end{sideways}
& \begin{sideways}Gly\end{sideways}
  \end{tabular}
  \caption{Molecular structure matrix. The entry in row $i$ and column
    $j$ denotes the number of steps required to transform the $i$th
    amino acid into the $j$th. (The gray tones are for
    ease of reading only, they do not carry special meaning.) \label{tab:molstructure} }
\end{table}

In order to perform the error-robustness calculations, we followed the
procedure by Haig and Hurst~\cite{haighurst91} and considered the
squared distances. In this way, the zeroes in the diagonal remain zero. The 
values for small changes become slightly larger (so the edge from Phe to Tyr 
gets a value 9) while the values for large changes (like going from Gly to Tyr) 
become considerably larger (in the case of Gly to Tyr 20 becomes 400). Large 
changes thus get stronger emphasis~\cite{digiulio89}. Whether squaring is the right way 
to make these kind of calculations has been discussed elsewhere~\cite{ardell98,freelandetal00}; we 
just want to compare molecular structure as an input to 
characteristics like polar requirement, hydropathy, volume and isoelectric point, as studied 
by~\cite{haighurst91}.The histograms of the error-robustness in terms of molecular structure 
are shown in the right column of Figure~\ref{fig:splithistograms}.

Although not producing (unlike polar requirement) a result in the less 
than 0.1 \% category, it is still remarkable that the SGC is,
with 0.151 \%, in the less than 1 \% category when molecular structure is used as input. This 
means that this matrix is performing better than volume or the hydropathy scale 
of hydrophobicity in the work of Haig and Hurst~\cite{haighurst91}.
Even more remarkable, the error robustness comes mainly from the second position, 
using this measure (Figure~\ref{fig:splithistograms}).

\subsection{Inverse Parametric Optimization} \label{app:eppstein}

Instead of asking the question ``What is the most error-robust genetic code in terms of e.g. polar requirement?'', 
one could also ask the question ``Is there a set of numerical values
for the 20 amino-acids such that the SGC is the optimal code in terms
of error robustness?'' If one particular set of 20 values turns out to have that property, one can compare 
this set with different sets of amino-acid characteristics, and suggest which characteristic resembles the 
``ideal values'' best. This then might be the factor playing a selective role during evolution of the SGC.

Let $A$ be the set of amino acids and let $\mathcal{F}$ be the set of
all codes. We aim at solving the following problem: Find a
non-trivial vector $\mathbf{x} \in \R_{\geq 0}^{A}$ of amino acid property values such that $\MS^{\alpha, \mathbf{x}}_0(\SGC) =
\argmin\limits_{F \in \mathcal{F}}\MS^{\alpha, \mathbf{x}}_0(F)$.

To solve this problem, we used a modification of the method of
Eppstein~\cite{eppstein03}. We define variables $\mathbf{x} \in \R^{20}$ and
consider the following constraint satisfaction problem: Find
$\mathbf{x}$ such that
\begin{align}
\mathbf{x} &\neq \mathbf{0}\label{eq:2}\\
\mathbf{x} &\geq \mathbf{0}\label{eq:3}\\
\MS^{\alpha,\mathbf{x}}_0(\SGC) &\leq \MS^{\alpha, \mathbf{x}}_0(F) & \text{ for all }
F \in \mathcal{F} \label{eq:1}
\end{align}

Note that the number of inequalities~\eqref{eq:1} equals the size of the
code space, which can be quite large. To deal with the potentially
large number of constraints we follow a \emph{cutting plane
  approach}. We work with intermediate solutions
$\overline{\mathbf{x}}_i$, start with $i = 0$, and set $\barx_0$ to some
random values that satisfy constraints~\eqref{eq:2} and
\eqref{eq:3}. We then solve the \emph{separation problem}
for the class of constraints~\eqref{eq:1}. That is, we have to find a code $F$
such that $\MS^{\alpha,\barx_i}_0(F) < \MS^{\alpha,\barx_i}_0(\SGC)$ or
prove that no such code exists. We can answer this question by finding
\[
F^* = \argmin_{F \in \mathcal{F}} \MS^{\alpha,\barx_i}_0(F)\enspace,
\]
using the quadratic assignment approach described in~\cite{buhrmanetal11}. 
In fact, for the actual procedure it suffices to
use much faster QAP heuristics, e.g., based on simulated
annealing~\cite{burkardrendl84} or the GRASP heuristic~\cite{lietal94}, instead of full QAP solvers.
If we find an $F$ with $\MS^{\alpha,\barx_i}_0(F) <
\MS^{\alpha,\barx_i}_0(\SGC)$, we have found a violated inequality 
\[
\MS^{\alpha,\mathbf{x}}_0(\SGC) \leq \MS^{\alpha,\mathbf{x}}_0(F)\enspace,
\]
which we add to the constraint satisfaction problem. We solve this set
of quadratic constraints using
the non-linear constraint solver \emph{fmincon} from MATLAB's
optimization toolbox~\cite{MATLAB11}, obtain a new set of values
$\barx_{i+1}$ and iterate the process until no more violated inequalities
can be separated. A final solution $\mathbf{x^*}$ can be verified by a QAP
solver such as~\cite{burkardderigs80}. All software used is provided
as supplemental information.

\begin{figure}[hbtp]
   \includegraphics[width=0.95\linewidth]{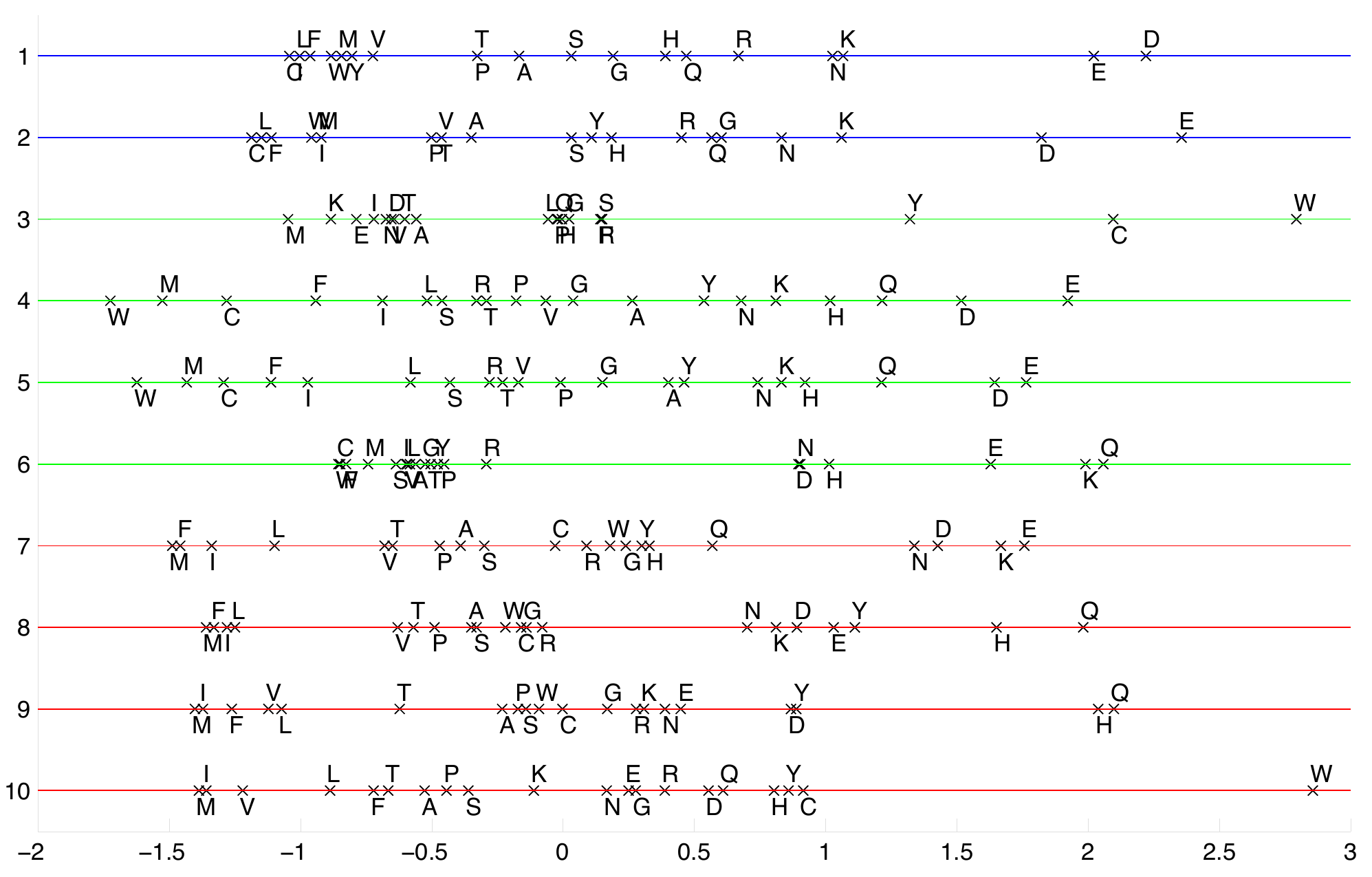}
\caption{Eight examples of sets of values for the 20 amino-acids that make the SGC the most error-robust genetic code. The (artificial) values are found by using inverse parametric optimization as described in Appendix~\ref{app:eppstein}. All sets have been normalized to have mean 0 and standard deviation 1. For comparison, we also show the original polar requirements on top (1), and the updated polar-requirement values on the second row (2).
Value sets 3 to 6 make the SGC optimal with respect to $\MS_0$. Value sets
7 to 10 make the SGC optimal with respect to $\MS_0^{\mathit{FH}}$.  \label{fig:Eppstein}}\end{figure}

Using this procedure, we found many different sets of 20 values under which
the SGC is optimal with respect to error-robustness. We steered the
values towards the polar requirement values $\mathbf{r}$ by using the
distance to $\mathbf{r}$ as the objective function in our approach. See
Figure~\ref{fig:Eppstein} for an illustration of some of the solutions
we found.

An analysis of the correlation coefficients of these ``ideal'' values
with a database of 744 known amino-acid properties from the literature
(AAindex:~\cite{kawashimaetal99}) shows no correlation above $0.82$
except with polar requirement. In other words, we do not know of any sets
of straightforward physico-chemical amino-acid properties which resemble 
one of these ``ideal'' sets.  This might suggest that a combination of 
several aspects of code evolution and amino-acid properties (as suggested by 
e.g.\ Higgs~\cite{higgs09}) resulted in the configuration of the SGC.

\subsection{Scan of Other Amino-Acid Properties}
\label{app:aaindex}

We performed error-robustness calculations for all (complete)
amino-acid properties of the
AAindex-database~\cite{kawashimaetal99}. For the purpose of comparison, we
extended the database to include the original polar requirements~\cite{woeseetal66a}, and the
updated polar requirements~\cite{mathewlutheyschulten08}, as well as
two sets of numerical values found by the procedure described in~\ref{app:eppstein}.

In a first scan, 50000 random codes were sampled from
\begin{enumerate}
\item all codes,
\item codes with the 7 assignments of Phe, Tyr, Trp, His,
  Leu, Ile, and Arg fixed,
\item codes with 7 fixed assignments and respecting the 
  structure enforced by the constraint of the historically 
  reasonable set of possible codes (all 11520 codes were 
  computed in this case).
\end{enumerate}
For all of the three settings above, error-robustness values were
computed using Haig-Hurst and Freeland-Hurst weights (the same
random samples were used for the two weight sets, the results are thus
statistically correlated).

Out of the 55 best-performing codes, the same calculations as above were
performed with $10^{6}$ samples. The 20 best performing properties are presented in
Table~\ref{tab:aaindex}. Not surprisingly, our two sets of
(artificial) numerical values found by inverse parametric optimization
(described in Appendix~\ref{app:eppstein}) end up on the top. 

Furthermore, we observe that the SGC is error-robust in terms of
several measures of polar requirement (as noted, e.g.,
in~\cite{vetsigianetal06}). One of these (for which this is not 
immediately obvious) is Grantham's polarity scale~\cite{grantham74}, 
which is a combination of Aboderin's scale~\cite{aboderin71} and 
polar requirement. It is especially noteworthy that the 
updated polar requirement~\cite{mathewlutheyschulten08} is 
consistently showing up within the best four sets of numerical 
values. When the sets found by inverse parametric optimization 
are left out, the updated values of polar requirement are in 
all three settings (no blocks fixed, 7 blocks fixed, and the 
set of 11520 codes resulting from 7 fixed blocks plus the 
constraint of the historically reasonable set of possible 
codes) the best set of values when Freeland-Hurst weights 
are used.

\begin{table}[h]
\centering
\tiny
\setlength{\tabcolsep}{0.8mm}
\begin{tabular}{|r l| r l| r l | r l | r l | r l | p{5cm}|}
\multicolumn{4}{|c|}{\tiny $10^6$ random codes} &
\multicolumn{4}{|c|}{\tiny $10^6$ random codes} & 
\multicolumn{4}{|c|}{\tiny 11520 codes}\\
\multicolumn{4}{|c|}{\tiny no blocks fixed} &
\multicolumn{4}{|c|}{\tiny 7 blocks fixed} & 
\multicolumn{4}{|c|}{\tiny 7 fixed, subsets}\\
\multicolumn{2}{|c}{HH} & \multicolumn{2}{|c|}{FH} &
\multicolumn{2}{|c}{HH} & \multicolumn{2}{|c|}{FH} &
\multicolumn{2}{|c}{HH} & \multicolumn{2}{|c|}{FH} & Description \\ 
\hline
0 & (1) &0 & (1) &0 & (1) &2 & (3) &0 & (1) &2 & (26) &{\it Some set of 20 values that make SGC optimal with Haig-Hurst weights (this study)}\\
\rowcol
1 & (2) &0 & (1) &1 & (2) &0 & (1) &0 & (1) &0 & (1) &{\it Some set of 20 values that make SGC optimal with Freeland-Hurst weights (this study)}\\
10 & (3) &4 & (6) &443 & (30) &13 & (10) &1 & (11) &3 & (33) &Long range non-bonded energy per atom (Oobatake-Ooi, 1977)~\cite{OobatakeO77}\\
\rowcol
17 & (4) &0 & (1) &6 & (3) &0 & (1) &0 & (1) &0 & (1) &{\it Updated Polar Requirements (Mathew, Luthey-Schulten 2008)}~\cite{mathewlutheyschulten08}\\
24 & (5) &40 & (16) &48 & (10) &21 & (13) &2 & (17) &0 & (1) &Information value for accessibility; average fraction 23\% (Biou et al., 1988)~\cite{BiouGLRG88}\\
\rowcol
30 & (6) &6 & (7) &26 & (5) &6 & (8) &6 & (35) &3 & (33) &Polarity (Grantham, 1974)~\cite{grantham74}\\
35 & (7) &57 & (18) &313 & (22) &44 & (18) &4 & (30) &5 & (44) &Free energies of transfer of AcWl-X-LL peptides from bilayer interface to \\
\rowcol
40 & (8) &130 & (31) &37 & (7) &27 & (15) &3 & (25) &0 & (1) &Surface composition of amino acids in intracellular proteins of mesophiles \\
46 & (9) &57 & (18) &205 & (21) &111 & (22) &0 & (1) &0 & (1) &Optimized relative partition energies - method D (Miyazawa-Jernigan, 1999)~\cite{MiyazawaJ99}\\
\rowcol
51 & (10) &26 & (11) &185 & (20) &8 & (9) &13 & (41) &1 & (19) &Effective partition energy (Miyazawa-Jernigan, 1985)~\cite{MiyazawaJ85}\\
58 & (11) &42 & (17) &55 & (12) &500 & (39) &1 & (11) &3 & (33) &Average side chain orientation angle (Meirovitch et al., 1980)~\cite{MeirovitchRS80}\\
\rowcol
96 & (12) &12 & (8) &173 & (19) &101 & (21) &3 & (25) &1 & (19) &Linker propensity from small dataset (linker length is less than six \\
98 & (13) &58 & (20) &623 & (37) &135 & (24) &32 & (50) &3 & (33) &Optimized relative partition energies - method C (Miyazawa-Jernigan, 1999)~\cite{MiyazawaJ99}\\
\rowcol
108 & (14) &34 & (13) &322 & (23) &3 & (5) &21 & (46) &4 & (40) &Optimal matching hydrophobicity (Sweet-Eisenberg, 1983)~\cite{SweetE83}\\
112 & (15) &37 & (14) &330 & (24) &3 & (5) &21 & (46) &4 & (40) &SWEIG index (Cornette et al., 1987)~\cite{CornetteCMSBL87}\\
\rowcol
119 & (16) &3 & (5) &41 & (8) &4 & (7) &2 & (17) &2 & (26) &Original Polar Requirements (Woese 1966)~\cite{woeseetal66a}\\
127 & (17) &23 & (10) &109 & (16) &38 & (17) &5 & (34) &1 & (19) &Average gain ratio in surrounding hydrophobicity (Ponnuswamy et al., 1980)~\cite{PonnuswamyPM80}\\
\rowcol
136 & (18) &1 & (4) &28 & (6) &2 & (3) &2 & (17) &2 & (26) &Polar requirement (Woese, 1973)~\cite{Woese73}\\
218 & (19) &95 & (28) &452 & (31) &235 & (31) &1 & (11) &0 & (1) &Information value for accessibility; average fraction 35\% (Biou et al., 1988)~\cite{BiouGLRG88}\\
\rowcol
279 & (20) &16 & (9) &120 & (17) &286 & (35) &2 & (17) &2 & (26) &Direction of hydrophobic moment (Eisenberg-McLachlan, 1986)~\cite{EisenbergL86}\\
\end{tabular}
  \caption{Table of the 20 most error-robust amino acid properties from the
    AAindex-database~\cite{kawashimaetal99}. The numbers indicate how
    many codes were found that are strictly more error-robust than the standard
    genetic code. The numbers in parantheses denote the rank among
    the 55 properties that have been analyzed. Description in {\it
      italic} indicate that this property is not included in the
    AAindex-database, but has been added for comparison. HH=Haig-Hurst, FH=Freeland-Hurst.\label{tab:aaindex} }
\end{table}

\subsection{Minimal Number of Fixed Assignments}
\label{app:minfixed}
In this appendix, we investigate how many amino-acid assignments need
to be fixed such that the SGC is the most error-robust genetic code
with respect to the updated polar requirements~\cite{mathewlutheyschulten08}, 
when we do \emph{not} use the constraint of the historically reasonable set 
of possible codes.

For the case of the Haig-Hurst weights, there are 67 different minimal
subsets $S_1, S_2, \ldots$, $S_{67} \subseteq \{\mathrm{Phe}, \mathrm{Leu}, \mathrm{Ile}, \ldots, \mathrm{Ser}, \mathrm{Gly}\}$
such that for any $i \in \{1,2,\ldots,67\}$, fixing the assignments of
all amino acids in $S_i$ makes the SGC the most error-robust genetic
code. Any super-set of these 67 minimal subsets will also have
this property, because fixing more assignments only limits the number
of possible genetic codes. Out of the 67 minimal subsets, 34 of them are of size 9, 15 of size 10, 15 of
size 11, and 3 of size 12. 

When fixing the seven assignments of Phe, Tyr, Trp, His, Leu, Ile, and
Arg (based on aptamer experiments) the minimal sets of assignments
that need to be fixed in addition are:
$\{\mathrm{Ser},\mathrm{Gln},\mathrm{Cys}\}$ or
$\{\mathrm{Met},\mathrm{Ser},\mathrm{Gln}\}$.

For the case of the Freeland-Hurst weights, there are 186 different minimal
subsets: 2 subsets of size 10, 4 of size 11, 13 of size 12, 44 of
size 13, 52 of size 14, 45 of size 15, 21 of size 16, and 5 of size 17. 
When fixing the seven assignments of Phe, Tyr, Trp, His, Leu, Ile, and
Arg (based on aptamer experiments), there are 6 different minimal sets
(of size 6) each of which can be fixed in addition in order to make
the SGC the most error-robust genetic code.

\end{document}